\documentclass[a4paper,11pt]{article}
\pdfoutput=1 

\usepackage{jinstpub} 
\usepackage{graphicx}
\usepackage{caption}
\usepackage{subcaption}
\usepackage{amsmath}   
\usepackage{xspace}

\usepackage[printwatermark]{xwatermark}
\usepackage{xcolor}
\usepackage{lipsum}
\usepackage{tikz}
\usepackage{lineno}
\usepackage{lipsum}
\usepackage{soul}

\title{Triple GEM performance in magnetic field}

\author[a]{M. Alexeev, }
\author[a,b]{A. Amoroso, }
\author[a]{S. Bagnasco, }
\author[c,d]{R. Baldini Ferroli, }
\author[c,e]{I. Balossino, }
\author[d]{G. Bencivenni, }
\author[d]{M. Bertani, }
\author[e]{D. Bettoni, }
\author[a,b]{F. Bianchi, }
\author[a,b]{A. Bortone, }
\author[d]{A. Calcaterra, }
\author[d,k]{M. Capodiferro, }
\author[e]{V. Carassiti, }
\author[d]{S. Cerioni, }
\author[a,d,f]{J. Chai, }
\author[a]{W. Cheng, }
\author[e]{S. Chiozzi, }
\author[e]{G. Cibinetto, }
\author[e]{A. Cotta Ramusino, }
\author[a,b]{G. Cotto, }
\author[a,f]{F. Cossio, }
\author[a]{M. Da Rocha Rolo, }
\author[a,b]{F. De Mori, }
\author[a,b]{M. Destefanis, }
\author[d]{D. Domenici, }
\author[c,d]{J. Dong, }
\author[e]{F. Evangelisti, }
\author[e,g]{R. Farinelli \note{Corresponding author.}, }
\author[a]{L. Fava, }
\author[d]{G. Felici, }
\author[e]{E. Fioravanti, }
\author[a]{L. Gaido, }
\author[e,g]{I. Garzia, }
\author[d]{M. Gatta, }
\author[a]{G. Giraudo, }
\author[a,b]{M. Greco, }
\author[a,c]{L. Lavezzi, }
\author[a]{C. Leng, }
\author[a]{H. Li, }
\author[h]{P. Li, }
\author[a]{S. Lusso, }
\author[a,b]{M. Maggiora, }
\author[e]{R. Malaguti, }
\author[i,l]{A. Mangoni, }
\author[a,b]{S. Marcello, }
\author[e]{M. Melchiorri, }
\author[e]{G. Mezzadri, }
\author[d]{G. Morello, }
\author[a]{M. Mignone, }
\author[c]{E. Pace, }
\author[i,l]{S. Pacetti, }
\author[d]{G. Papalino, }
\author[a,b]{B. Passalacqua, }
\author[d]{P. Patteri, }
\author[d,m]{A. Pelosi, }
\author[d]{M. Poli Lener, }
\author[a]{A. Rivetti, }
\author[e,g]{M. Savri\'e, }
\author[e,g]{M. Scodeggio, }
\author[a]{S. Sosio, }
\author[a,b]{S. Spataro, }
\author[d,n]{E. Tskhadadze, }
\author[e,g]{S. Verma, }
\author[a]{L. Yang, }
\author[c]{B. Wang, }
\author[a]{R. J. Weadon, }
\author[c]{J. Zhang.}

\affiliation[a]{INFN, Sezione di Torino, via P. Giuria 1, 10125 Torino, Italy}
\affiliation[b]{Universit\`a di Torino, Dipartimento di Fisica, via P. Giuria 1, 10125 Torino, Italy}
\affiliation[c]{Institute of High Energy Physics, Chinese Academy of Sciences, 19B YuquanLu, Beijing, 100049, People's Republic of China}
\affiliation[d]{INFN, Laboratori Nazionali di Frascati, via E. Fermi 40, 00044 Frascati (Roma), Italy}
\affiliation[e]{INFN, Sezione di Ferrara, via G. Saragat 1, 44122 Ferrara, Italy}
\affiliation[f]{Politecnico di Torino, Dipartimento di Elettronica e Telecomunicazioni, Corso Duca degli Abruzzi 24, 10129 Torino, Italy}
\affiliation[g]{Universit\`a di Ferrara, Dipartimento di Fisica, via G. Saragat 1, 44122 Ferrara, Italy}
\affiliation[h]{USTC, University of Science and Technology of China, Hefei 230026, People's Republic of China}
\affiliation[i]{INFN, Sezione di Perugia, via A. Pascoli, 06123 Perugia, Italy}
\affiliation[l]{Universit\`a di Perugia, Dipartimento di Fisica e Geologia, via A. Pascoli, 06123 Perugia, Italy}
\affiliation[m]{INFN Sezione di Roma1, I-00185, Roma, Italy}
\affiliation[n]{Technological Institute of Georgia ,Tbilisi, Georgia}

\emailAdd{rfarinelli@fe.infn.it}

\abstract{Performance of triple GEM prototypes in strong magnetic field has been evaluated by means of a muon beam at the H4 line of the SPS test area at CERN. Data have been reconstructed and analyzed offline with two reconstruction methods: the charge centroid and the micro-Time-Projection-Chamber exploiting the charge and the time measurement respectively. A combination of the two reconstruction methods is capable to guarantee a spatial resolution better than 150 $\mu$m in magnetic field up to a 1 T.}

\begin{document}
\maketitle
\flushbottom

\section{Introduction}

Gas Electron Multipliers (GEMs) have been invented by F. Sauli \cite{ref:sauli} in 1997 and in the past decades applications of this technology increased. In high energy physics GEMs are employed in high rate, high dose environments due to their very robust structure; they are used as tracking devices for their good spatial resolution, as in the case of KLOE-2  Cylindrical GEM detector \cite{ref:kloe}, or their time performance as in the LHCb muon chamber \cite{ref:LHCb} or the TOTEM experiment \cite{ref:totem}.

A GEM detector is composed by a thin (50 $\mu$m) Kapton foil, copper clad on each side, with a high surface density of holes \cite{ref:sauli}. In the standard technique each hole has a bi-conical structure with external (internal) diameter of 70 $\mu$m (50 $\mu$m); the hole pitch is 140 $\mu$m. The bi-conical shape of the hole is a consequence of the single mask process used in standard photolithographic technologies.
A typical voltage difference in a range between 300 to 500 V is applied between the two copper sides, producing an electric field reaching values up to 100 kV/cm into the holes. The effect of such high electric field is to multiply the number of the electrons produced by a charged particle crossing the detector by a factor up to a few thousands. Multiple structures realized by assembling two or more GEM foils at close distance allow to reach high gains, while minimizing the discharge probability \cite{ref:bachmann}.

So far few measurement of GEM detector performance with analog readout in a strong magnetic field has been published in literature. We performed several test beams (TB) with planar prototypes within the RD51 collaboration at CERN \cite{ref:rd51} to determine the achievable performance triple GEMs in magnetic field. 

We tested two reconstruction techniques. Charge Centroid method exploits the weighted mean to extract the track position at the anode. This method is widespread, not only in GEM technology, but it has never been tested extensively in strong magnetic field. Micro-TPC ($\mu$TPC) is an innovative readout, firstly developed for ATLAS MicroMegas\citep{ref:kostas}, that allows to reconstruct the position by means of the study of the time of arrival of the induced signals on the strips. 

In this paper we report results obtained with charge centroid and the first application of the $\mu$TPC reconstruction for a GEM detector in magnetic field up to 1~T.

\vspace{0.5cm}
\section{Test setup and analysis method}

The TBs have been performed at the H4 line of SPS test area at CERN, where a muon beam with features listed in Table \ref{tab:beam} \cite{ref:h4} is available.
The magnetic field is generated by a dipole magnet composed by two ferromagnetic disks having a diameter of 200 cm. The distance between the two magnetic poles is 106 cm. The magnet can provide a field strength up to 1.5 T in both polarizations. \\

\begin{table}[h]
\caption{Typical beam parameters during the BESIII Test Beam.}
\begin{center}
\begin{tabular}{cc}
\hline
Particles        &   $\mu^{+}$ and $\pi^{+}$ \\
Momentum    &  150 GeV/c \\
$\Delta$p/p 	              &  1.4\%   \\
Particles per spill &      [10$^{3}$ - 10$^{6}$] particles \\
Spill duration & $\sim$ 3 sec  \\
Duty cycle     & $\sim$  2 spills per minute \\
\hline
\end{tabular}
\label{tab:beam}
\end{center}
\end{table}

The experimental setup is composed of:

\begin{itemize}
\item Two $10\times10$ cm$^2$ triple GEM test chambers: one with $XY$ linear-strip readout, the other with $XV$ jagged-strip readout\footnote{A readout plane with jagged strips has been studied in order to lower the coupling capacitance between the two views; in this design the strip width is reduced in the overlap region.} \cite{ref:TIPP2014} and a stereo angle of $60^{\circ}$. The test chambers are placed inside the dipole magnet.
\item Two tracking stations, one 3.5 m upstream and one 3.5 m downstream w.r.t. the test prototypes; each station consists of two 6-cm-spaced $XY$ triple GEM chambers.
\item Scintillator detectors for the trigger: one upstream the forward trackers and one downstream the backward trackers. The trigger active area is about $4\times4$ cm$^2$.
\end{itemize}

\begin{figure}[tbp]
    \centering
        \includegraphics[width=0.6\textwidth]{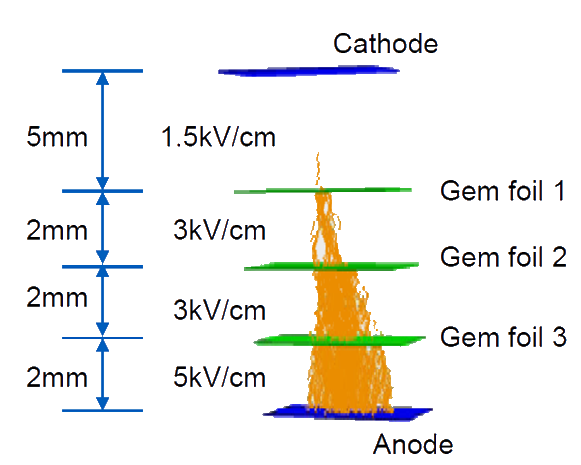}
        \caption{Representation of a triple GEM detector. Blue foils are the cathode and the anode and the green ones are the three amplification stages. On the left the applied field and the distances between each electrodes are shown. The Orange lines represents the path of the electrons with the opening of the avalanche inside the detector.}
        \label{fig:triple}
\end{figure}

In a triple GEM the five electrodes (cathode, three GEM foils and anode) define four gaps, namely: {\it conversion and drift}, between the cathode and the first GEM foil; {\it transfer 1} and {\it transfer 2}, between the GEM foils, and {\it induction} where the signal is induced on the readout plane. A representation of the internal structure is given in Fig. \ref{fig:triple}.
Standard GEM detectors have a 3 mm conversion gap and 2 mm for all the others (3/2/2/2). A configuration with wider conversion gap (5/2/2/2) has also been extensively studied, since it is expected to have better performance with the $\mu$TPC reconstruction. 
The strip pitch of the prototypes readout plane is 650 $\mu$m, the width of the $X$ strips (parallel to the magnetic field) is 130 $\mu$m, while the width of other coordinate strips is 570 $\mu$m ; the design has been optimized in order to have an equal sharing of the charge on the two views. The anode design is similar to the one used by the COMPASS experiment \cite{ref:compass}. 

\begin{figure}[htbp]
    \centering
    \begin{subfigure}[b]{0.48\textwidth}
        \includegraphics[width=\textwidth]{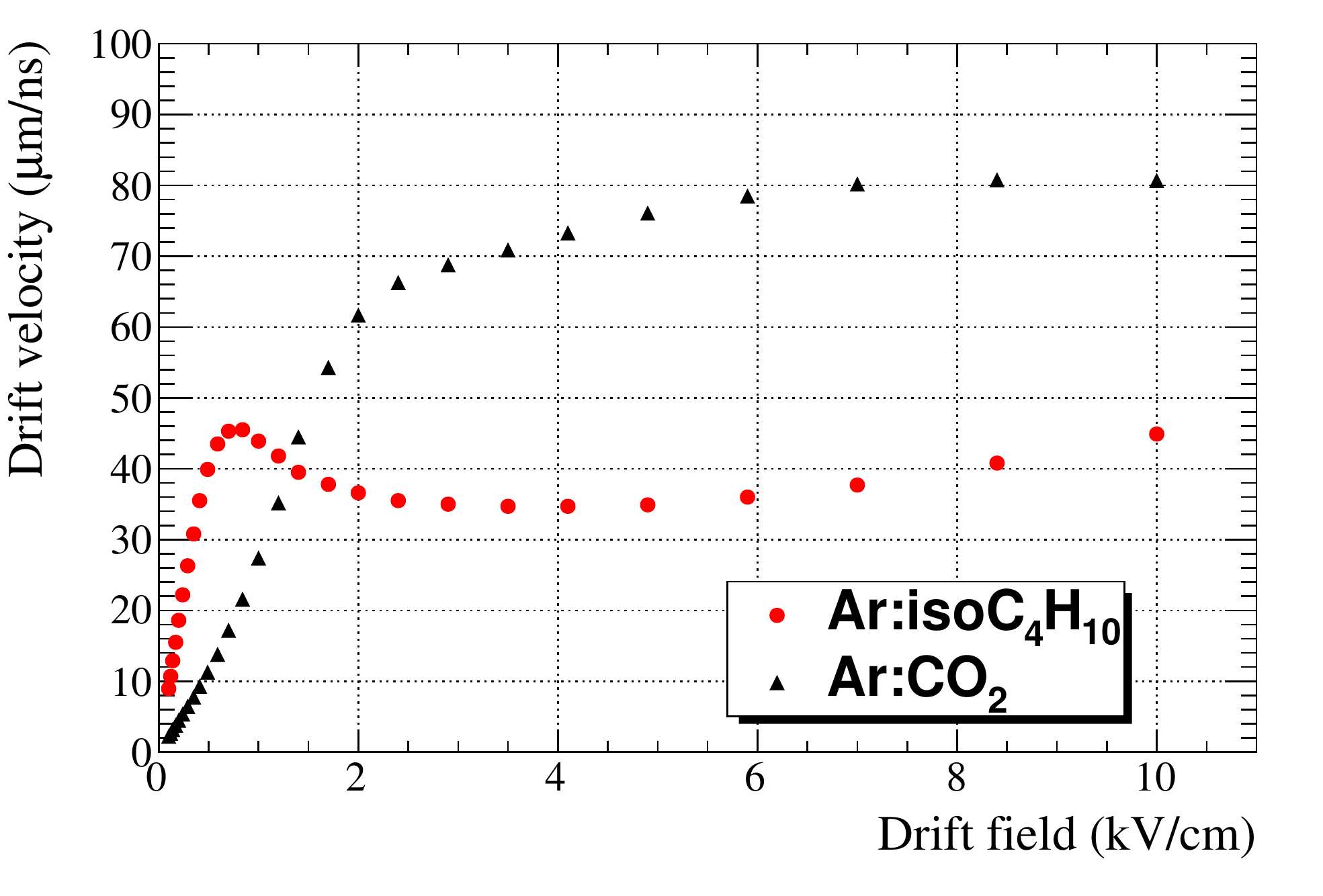}
        \caption{}
        \label{fig:DriftVelocity}
    \end{subfigure}
    ~ 
        \begin{subfigure}[b]{0.48\textwidth}
        \includegraphics[width=\textwidth]{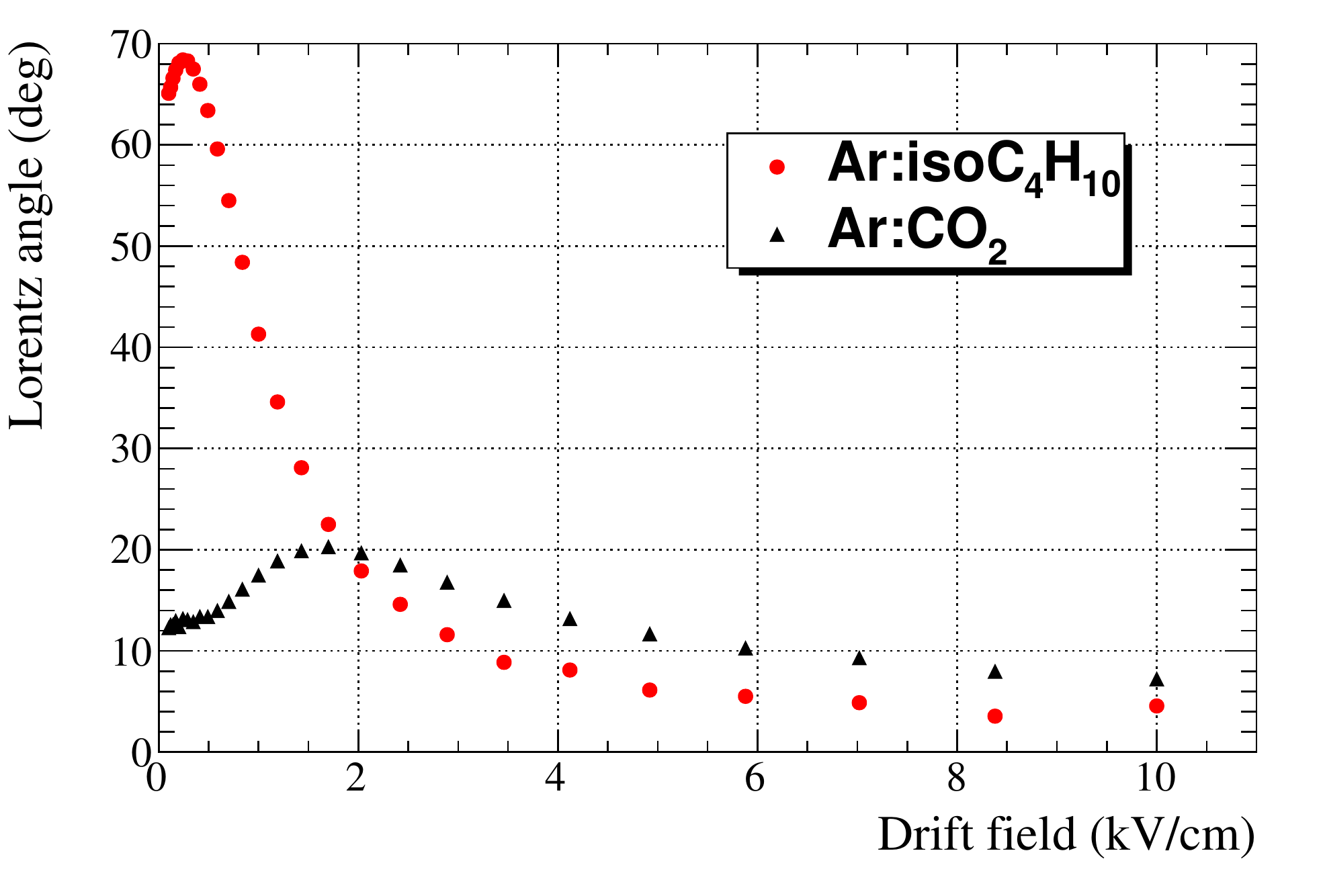}
        \caption{}
        \label{fig:LorentzAngle}
    \end{subfigure}
    \caption{Drift velocity (a) and Lorentz angle (b) Garfield simulations for B = 1 T.}
    \label{fig:Gas}
\end{figure}

The avalanche diffusion and the charge collection depend largely on the value of the electric fields within the gaps; based on our measurements and other R\&D studies \cite{ref:sauli2,ref:fieldset} we adopted the following standard electric field settings: 
\\
$E_{d}~=~1.5$~kV/cm in the conversion gap, $E_{t_1} = 3.0$ kV/cm, $E_{t_2} = 3.0$ kV/cm in the transfer gaps, $E_{i} = 5.0$ kV/cm in the induction gap in most gas mixture configuration. Unless differently stated, the sum of the voltages applied to the three GEM foils is set to provide an effective gas gain of about 9000 \cite{ref:sauli2}.
The test chambers are allowed to rotate on their vertical axis in order to acquire data with particles crossing the prototypes with different incident angles. This enable to study the behavior of non-orthogonal tracks. The X coordinate of the triple GEM is sensitive to the angle rotation and the magnetic field. The further results are focused on the X coordinate measurements.
All the GEM chambers are readout by the SRS system \cite{ref:srs} featuring APV25 hybrid cards, which provide an analog readout of the charge with 25 ns sampling time, allowing also the time measurement \cite{ref:apv25}. 
The gas system uses premixed gas bottles: Ar:CO$_2$ (70:30) and Ar:isoC$_4$H$_{10}$ (90:10) mixtures have been tested.
The effect of the magnetic field on the detector has been studied with Garfield \cite{ref:garfield} simulations in Ref. \cite{ref:rf_magistrale}. Drift velocity and Lorentz angle distributions as a function of the electric field have been evaluated with Magboltz \cite{ref:magboltz}, a software that solves the Boltzmann transport equations for electrons in gas mixtures under the influence of electric and magnetic fields. Where the Lorentz angle is the angle between the electric field and the electron drift direction. The results are reported in Fig. \ref{fig:Gas} for 1 T magnetic field. \\

For each event, a track is reconstructed by a linear fit of the clusters\footnote{A cluster is a group of contiguous strips fired at the same time.} on the tracking chambers outside the field region (one for each chamber). The expected position of the cluster on the test chambers is extracted by the interception of the reconstructed track and the test chambers plane; the good clusters in the test chamber are the one with the higher charge.
The residual distribution of the cluster position on the first test chamber w.r.t. the good cluster position on the other one is fitted with a Gaussian function. The extracted sigma divided by $\sqrt{2}$ is the extracted position resolution of the test chamber, assuming that the two have the same spatial resolution. This technique allows to remove the contribution of the tracking system and other source of systematic errors.

As counter check, the spatial resolution has been also measured as the standard deviation of the Gaussian fit to the residual distribution of the position measured by the test prototypes and the expected one estimated by the tracking system, as described above. This method subtract the contribution of the tracking system and the systematic errors to the residual distribution of the test chamber. To remove most of the systematic error due to the beam spread and to the tracking system a GEANT4 \cite{ref:geant} simulation has been used. The main contribution to the spatial resolution measurement is given by the beam momentum spread in magnetic field and it has been evaluated to be less than 100 $\mu$m at the positions of the test chambers. The resolution of the tracking chambers gives a contribution of the order of 50 $\mu$m. The effect of the alignment and fitting procedure is found to be negligible. The two methods give similar spatial resolution results. \\

The efficiency measurement of each prototype depends on the efficiency of the two prototypes system with respect to the tracking system. As described in Eq. \ref{eq:eff23}, events with at least one cluster in each of the four trackers have been selected. For the test chambers the number of events whose reconstructed position fall within 5 $\sigma$ of the residual distribution is calculated. The efficiency of both test chambers is the ratio between the number of events within 5 $\sigma$ divided by the number of selected events.The residual is measured comparing the position of the two prototypes, this is the reason why an efficiency of both chamber together is evaluated at first.
\begin{equation} 
\epsilon_{proto_1\&proto_2} = \frac{\#\>events\>within\>5\>\sigma}{\#\>tracked \>events}
\label{eq:eff23}
\end{equation}
Assuming the same performance for the two prototypes, the square root of $\epsilon_{proto_1\&proto_2}$ is used to measure the efficiency of a single chamber as described in Eq. \ref{eq:eff}. This approach underestimates (overestimates) the efficiency of the chamber with 5 (3) mm conversion gap because higher efficiency is expected when the drift gap is larger. \\
\begin{equation} 
\epsilon_{proto_1} = \epsilon_{proto_2} = \epsilon = \sqrt{\epsilon_{proto_1\&proto_2}}
\label{eq:eff}
\end{equation}

\vspace{0.5cm}
\section{Results with charge centroid method}

The charge centroid (CC) or centre of gravity method exploits the charge distribution of the strip clusters to improve the spatial resolution with respect to the digital readout, where the resolution is the cluster size divided by $\sqrt{12}$. The CC uses contiguous fired strips and a weighted average is performed with the charge measured by each strip and their position. 
Assuming a Gaussian avalanche profile, the reconstructed shape of the charge distribution at the anode allows to improve the spatial resolution as long as the cluster multiplicity\footnote{The number of adjacent strips composing a cluster.} is greater than one and the electrical signal shape is symmetric. \\

\begin{figure}[htbp]
    \centering
    \begin{subfigure}[b]{0.48\textwidth}
        \includegraphics[width=\textwidth]{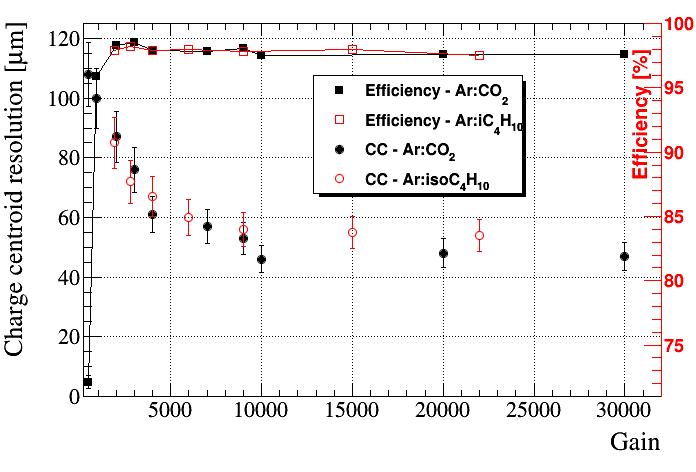}
       \caption{}
        \label{fig:EffVsGain}
    \end{subfigure}
    ~
        \begin{subfigure}[b]{0.48\textwidth}
        \includegraphics[width=\textwidth]{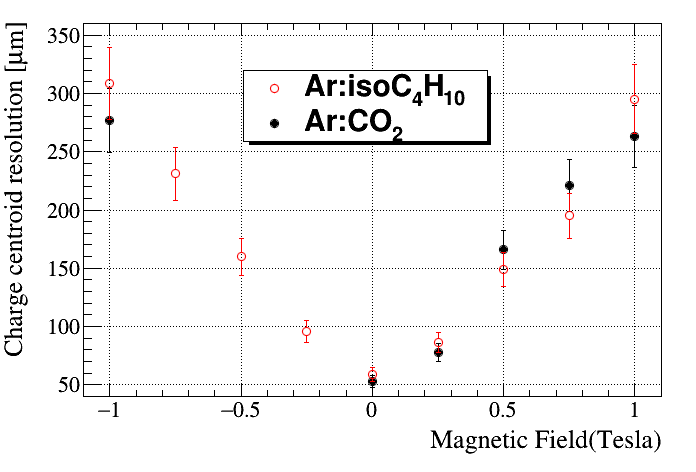}
        \caption{}
        \label{fig:ResVsB}
    \end{subfigure}
    \caption{Performance of a triple GEM prototype with charge centroid for Ar:CO$_2$ (70:30) and Ar:isoC$_4$H$_{10}$ (90:10) gas mixtures. The conversion gap of the prototype is 5 mm. (a) Detection efficiency and charge centroid resolution for B = 0 T. Results obtained with a drift field of 1.5 kV/cm. (b) Spatial resolution as function of the magnetic field intensity (Gain $\sim$ 9000).}\label{fig:CC1}
\end{figure}

For particles crossing the detector orthogonally and without magnetic field the charge distribution depends mainly on the electron diffusion within the gaps, that is driven by the gas mixture composition and the applied electric fields. Under these conditions, both detection efficiency and spatial resolution are very stable for different gas mixtures and operating parameters. The mean value of one dimensional efficiency on the plateau region is above 98\% and the spatial resolution is below 50 $\mu$m and depends mainly on the cluster hit multiplicity. The efficiency plateau begins at a gain of about 5000 as shown in Fig. \ref{fig:CC1}a \cite{ref:nimCGEM}.\\

When a strong magnetic field orthogonal to the electron drift direction is applied the Lorentz force displaces the electronic cloud; the effect is a heavy smearing and deformation of the charge distribution shape at the anode. The signal is no more Gaussian,  thus not easy to be parametrized: the stronger is the Lorentz force, the wider is the smearing. Therefore the spatial resolution of the CC method starts degrading linearly with the intensity of the magnetic field as shown in Fig. \ref{fig:ResVsB}. On the other hand, the detection efficiency is not affected and remains as good as it is without magnetic field.\\

\begin{figure}[htbp]
    \centering
        \includegraphics[width=0.6\textwidth]{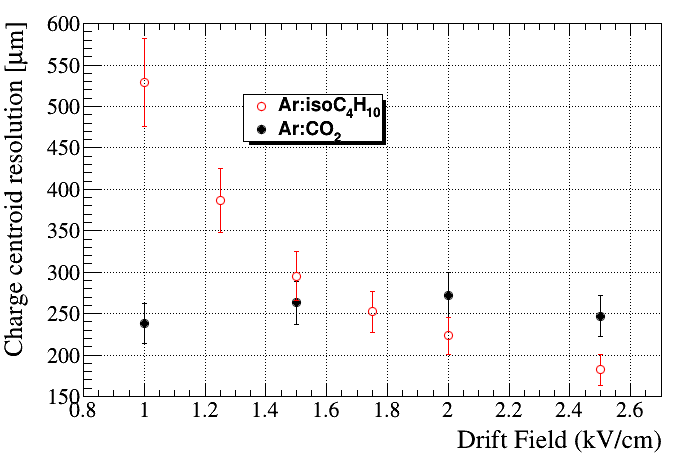}
        \put(-85,80){B = 1 T}
        \caption{Triple GEM spatial resolution as a function of the drift field for different gas mixtures. Results obtained with
        a gas gain of 9000.}
        \label{fig:ResVsDrift}
\end{figure}

Since most of the diffusion occurs in the conversion gap, that is thicker and has lower electric field compared to the other gaps, we performed a scan on the drift field to minimize the Lorentz effect and optimize the high voltage settings for operations in a "strong magnetic field"; the study has been performed at B = 1 T, with a gas gain of 9000 for prototypes with different conversion gaps and different gas mixtures. \\
The results are reported in Fig. \ref{fig:ResVsDrift}. The spatial resolution dependency on the drift field has the same behaviour as the Lorentz angle, which is shown in Fig. \ref{fig:LorentzAngle}. Using Ar:isoC$_4$H$_{10}$ (90:10) gas mixture and 3 mm drift gap we achieved the unprecedented resolution, for a GEM detector in a 1 T magnetic field, of about 180 $\mu$m. \\

The average size of the clusters ranges from four to eight strips, depending on the chamber settings, therefore a finer strip pitch does not help to further improve the space resolution for the CC method. Improvements of the order of 10-30 $\mu$m can be achieved by means of better clustering algorithms ({\it i.e.} removing the tails of the big clusters, taking into account the charge saturation, etc.). The efficiency goodness and stability were checked within the drift field range under study. \\

The best results achieved with the CC in magnetic field is 180 $\mu$m. To reach better performance an alternative method has to be taken into account. 

\vspace{0.5cm}
\section{Development and results of the $\mu$TPC reconstruction}
\label{sec:TPC}

In order to improve the spatial resolution of GEM detectors in magnetic field and with inclined tracks, it is possible to exploit the single strip time information to perform a local track reconstruction in the few mm drift gap like in a small Time Projection Chamber. This reconstruction method is called micro-Time-Projection-Chamber ($\mu$TPC) mode and can be efficient only if the time resolution of the detector and the front end electronic are capable to distinguish the arrival time of the electron avalanche on the different strips. The position of each strip provides one of the two required coordinates, while the other coordinate z (perpendicular to the strip plane) is reconstructed from the time measurement and the electron drift velocity extracted from Garfield simulations, as described in Eq. $\ref{eq:tpc}$ . v$_{drift}$ is the drift velocity, t is the measured time and $\overline{t}$ is drift time of an electron from the first GEM plane to the anode. $\overline{t}$ is approximately constant for each electron. Only the time taken by the primary electron from the place where is generated to the first GEM is needed. 

\begin{equation} 
z_{\mu TPC} = v_{drift} \times \left( t-\overline{t} \right)
\label{eq:tpc}
\end{equation}

The $\mu$TPC reconstruction technique has been originally developed by the ATLAS MicroMegas collaboration \cite{ref:atlasTPC}\cite{ref:atlasTPC2}.\\

\begin{figure}[tbp]s
    \centering
        \includegraphics[width=0.6\textwidth]{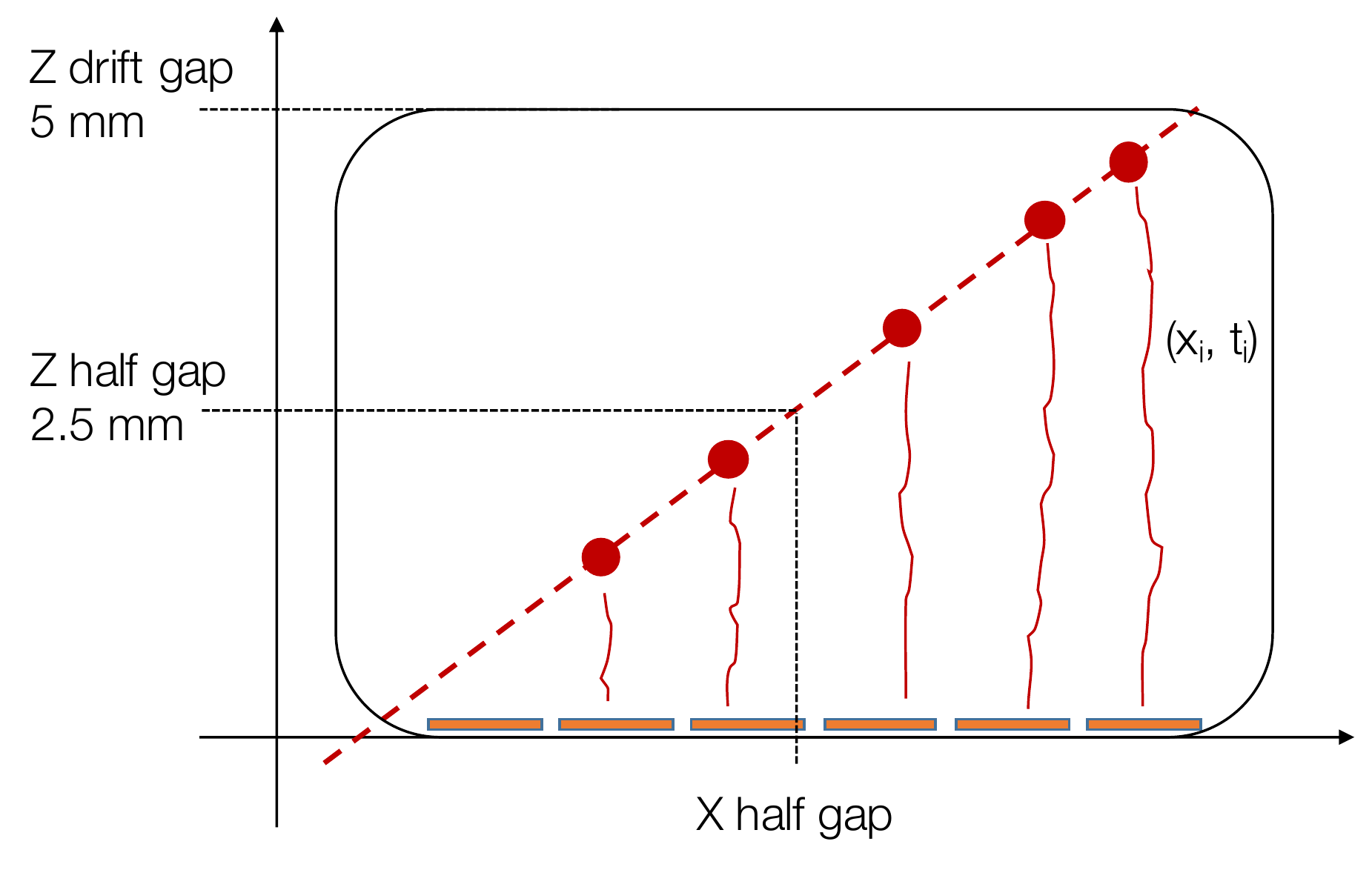}
        \caption{Sketch of the track reconstruction inside the conversion gap with the $\mu$TPC method.}
        \label{fig:uTPC}
\end{figure}

In Fig. \ref{fig:uTPC} the $\mu$TPC concept is explained. Once a cluster is found, coordinates ($x,z$) and errors ($dx, dz$) are assigned to each strip and a fit with a straight line is performed. Errors $dx$ basically account for the uncertainty of the hit in the finite strip pitch plus a weight depending on the fraction of the total charge collected on the strip; $dz$ is the error as propagated from the time measurement error. 

The $\mu$TPC clustering method has been initially tested with inclined tracks and no magnetic field; data with chambers at 10$^\circ$, 20$^\circ$, 30$^\circ$ and 45$^\circ$ w.r.t. the beam direction have been collected. An angular bias is observed in the reconstructed values of the angles measured by the $\mu$TPC. The effect has been understood to be due to the capacitive induction \citep{ref:kostas} of the signal on neighboring strips (the first and last strips in the strip cluster playing the major role). Induced signals in the first and last strips of a cluster have same arrival times as their neighbours, and in the $\mu$TPC reconstruction the effect is a tilt of the original track towards larger angles. In addition the transverse diffusion plays another important role since the signal of a single primary electron can be spread on more than one strip. A data driven correction procedure has been studied to take into account this effect. The induced signal is identified studying the time information and charge ratio of the strip w.r.t. its neighbour. A subsequent weighting or suppression of the first and/or last strips of the cluster has been implemented. The correction algorithm is still in a preliminary stage, but first results showed a reduction of the angle bias. \\

\begin{figure}[tbp]
    \centering
        \includegraphics[width=0.6\textwidth]{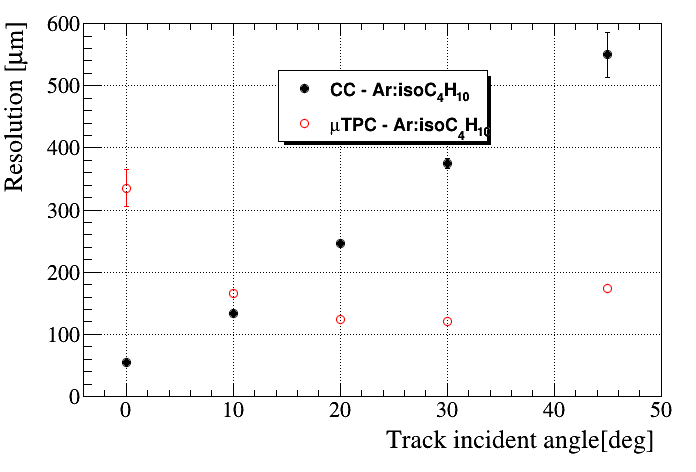}
        \put(-170,125){B = 0 T}
        \caption{Spatial resolution of the CC and $\mu$TPC cluster reconstruction vs the incident angle of the track for Ar:isoC$_4$H$_{10}$ (90:10) gas mixture at B = 0 T. Results obtained with a drift gap of 5 mm, a drift field of 1.5 kV/cm and a gain of 9000.}
        \label{fig:Res_vs_angle_B0T}
\end{figure}

The spatial resolution has been calculated as the difference of the reconstructed position of the two prototypes, both reconstructed in $\mu$TPC mode. The residual distribution can be fitted by a double Gaussian function with a narrow core contribution and a wider contribution that takes into account anomalous clusters. The spatial resolution is computed as the weighted average of the two Gaussian widths and assuming, as before, the same resolution for the two chambers, except for a geometric factor. This accounts for the different thicknesses of the conversion gaps in the two prototypes and can be computed as the square root of their ratio \cite{ref:kostas}. Fig. \ref{fig:Res_vs_angle_B0T} shows the distribution of the resolution for a 5 mm gap prototype with Ar:isoC$_4$H$_{10}$ (90:10) gas mixture as a function of the incident angle, in comparison with the CC one: spatial resolution better than 200 $\mu$m is achievable for impact angles up to 45$^\circ$. 
The $\mu$TPC method is limited by the time resolution, that is the convolution of the detector and electronics contributions, both being about 8 ns. 

\begin{figure}[bb]
    \centering
        \includegraphics[width=0.8\textwidth]{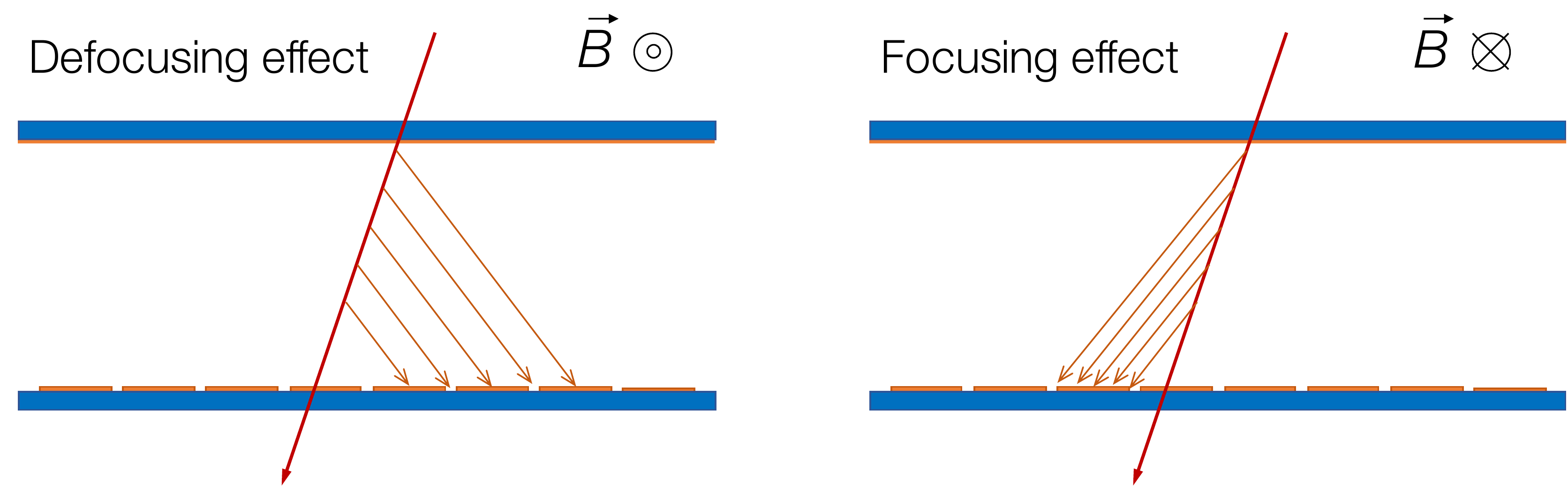}
        \caption{Defocusing and focusing effect on a general MPGD.}
        \label{fig:focus}
\end{figure}

Fig. \ref{fig:focus} shows the effect of the magnetic field on the displacement of the avalanche in a general Multi Pattern Gas Detector (MPGD) for different tracks. Depending on the relative signs of the track angle $\vartheta_{track}$ and the Lorentz angle $\vartheta_{L}$ a $\textit{focusing}$ or $\textit{defocusing}$ effect is expected, where the former spreads the primary ionization over a smaller number of strips, the latter over a larger one. A $\textit{singular}$ configuration is reached when the particle track inclination is equal to the Lorentz angle $\vartheta_{track} = \vartheta_{L}$\footnote{This configuration has been labeled as "maximum focusing effect".}. From the point of view of track reconstruction, this condition is equivalent to an orthogonal track in absence of magnetic field, where the clusters have the minimal spread (the smallest size) and the CC provides the best spatial resolution. \\

\begin{figure}[h!]
    \centering
        \includegraphics[width=0.7\textwidth]{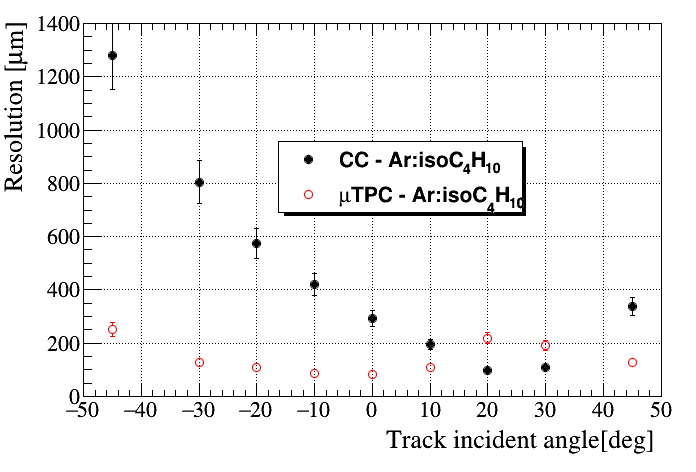}
        \caption{Spatial resolution for the CC (black dot) and $\mu$TPC (red circle) modes as a function of the incident angle of the track at B = 1 T.
        The data have been collected with Ar:isoC$_4$H$_{10}$ (90:10) gas mixture, and a gain of 9000. The conversion gap is 5 mm and the drift field is 1.5 kV/cm.}
        \label{fig:Res_vs_angle_B1T}
\end{figure}

From the Garfield simulation reported in Fig. \ref{fig:Gas}, for 1 T magnetic field, at a drift field of 1.5 kV/cm we expect a drift velocity of about 3.8 cm/$\mu$s and a Lorentz angle of $\sim26^\circ$. 
Fig. \ref{fig:Res_vs_angle_B1T} shows the results of the $\mu$TPC method as function of the incident angle in a 1 T magnetic field for a 5 mm gap chamber with Ar:isoC$_4$H$_{10}$ (90:10) gas mixture. As expected the $\mu$TPC mode performs well for negative angles ({\it i.e.} defocusing), while,  between +20$^\circ$ and +30$^\circ$, where  $\vartheta_{track} = \vartheta_{L}$, consistently with the Lorentz angle from the Garfield simulation, the CC method gives the best spatial resolution. For angles larger than 30$^\circ$ the focusing effect decreases and the $\mu$TPC algorithm becomes effective again. Negative angles are obtained by reversing the direction of the magnetic field. Moreover looking at the time distribution of the measured signals it is possible extract the time difference between the primary electrons generated close to the cathode and the one close to the first GEM. This time information together with the gap value allows to calculate the drift velocity from the data. Results agrees with the simulation within 10-15\%. 

\vspace{0.5cm}
\section{Discussion of the results}

For the first time, the performance of a triple GEM detector with analog readout has been evaluated in a strong magnetic field; a spatial resolution of 180 $\mu$m with the CC reconstruction, for orthogonal tracks at B = 1 T, has been measured using a high drift field (2.5 kV/cm) and Ar:isoC$_4$H$_{10}$ (90:10) gas mixture.  
For track incident angles larger than 10$^\circ$ the spatial resolution rapidly gets worsen. 
The CC performance depends mainly on the transverse diffusion which is affected by the Lorentz force in magnetic field, and by the drift gap size of the GEM detector.

A $\mu$TPC readout has been implemented and successfully tested for the first time for a GEM detector operated in magnetic field. 
The $\mu$TPC technique has been proved to be solid and well understood by the study with Garfield simulation. 

A combination of the two methods has been investigated to improve and to keep constant the spatial resolution determination for a wide range of incident track angles.

We tested Ar:CO$_2$ (70:30) and Ar:isoC$_4$H$_{10}$ (90:10) gas mixtures. With the CC readout the latter has a better spatial resolution at high drift fields ({\it i.e.} $E_{drift} \ge 1.5$~kV/cm) and the former has more solid performance. The results with CC and $\mu$TPC show a similar behavior with Ar:isoC$_4$H$_{10}$ and Ar:CO$_2$ gas mixtures.

Prototypes with 3 mm and 5 mm drift gap have been tested. For the CC method the 3 mm has a better spatial resolution with the Ar:isoC$_4$H$_{10}$ gas mixture while the 5 mm performs slightly better with Ar:CO$_2$. With the $\mu$TPC readout the 5 mm is certainly better since it can account for more points to reconstruct the track. 


\vspace{0.5cm}
\section{Conclusions}

GEM prototypes performance were evaluated with the analysis of data from a TB. The study showed that one-dimensional cluster efficiency is above 98\% for a wide range of operational settings. Concerning the spatial resolution, the analog readout performs extremely well with no magnetic field where the CC resolution is well below 100 $\mu$m. Increasing the magnetic field the resolution degrades linearly as an effect of the Lorentz force that displaces, broadens and produces asymmetries in the electron avalanche. Tuning the electric fields of the GEM prototype we achieved the unprecedented spatial resolution of 180 $\mu$m at 1 T. Space for improvements is still possible with the optimization of the clustering algorithms.\\

In order to improve the spatial resolution with strong magnetic field and inclined tracks a $\mu$TPC readout has been investigated. Such a readout mode exploits the good time resolution of the GEM detector and electronics to reconstruct the trajectory of the particle inside the conversion gap. Beside the improvement of the spatial resolution, information of the track angle can be also extracted. This is the first use of a $\mu$TPC readout with GEM detector in magnetic field.\\

The combination of the CC and $\mu$TPC reconstruction provides a flat spatial resolution around 100-120 $\mu$m at 1 T magnetic field for track incident angles ranging from -30$^\circ$ to 30$^\circ$. 


\vspace{0.5cm}
\acknowledgments

This work is supported by the Italian Institute of Nuclear Physics (INFN), by the MAE Executive Program PGR00136, and by the H2020-MSCA-RISE 2014 BESIIICGEM project. We thank the KLOE-2 collaboration for sharing their experience. We acknowledge the RD51 collaboration for their support, particularly Prof. Yorgos Tsipolitis and Dr. Eraldo Oliveri for their assistance during the TB. We thank Prof. Theodoros Alexopoulos and Dr. Konstantinos Ntekas for their suggestions on the $\mu$TPC readout.  \\


\end{document}